\documentclass[pra]{revtex4}
\usepackage{amssymb}
\usepackage{amsmath}
\usepackage{graphicx}

\begin{document}

\title{One- and two-dimensional solitons in spin-orbit-coupled Bose-Einstein
condensates with fractional kinetic energy}
\author{Hidetsugu Sakaguchi$^{1}$ and Boris A. Malomed$^{2,3}$}

\address{$^{1}$Department of Applied Science for Electronics and Materials, Interdisciplinary Graduate School of
Engineering Sciences, Kyushu University, Kasuga, Fukuoka 816-8580, Japan}
\address{$^{2}$Department of Physical Electronics, School of Electrical Engineering,
Faculty of Engineering, and Center for Light-Matter Interaction, Tel Aviv
University, P.O. Box 39040 Tel Aviv, Israel\\
$^{3}$Instituto de Alta Investigaci\'{o}n, Universidad de Tarapac\'{a}, Casilla 7D,
Arica, Chile}

\begin{abstract}
We address effects of spin-orbit coupling (SOC), phenomenologically added to
a two-component Bose-Einstein condensate composed of particles moving by L%
\'{e}vy flights, in one- and two-dimensional (1D) and (2D) settings. The
corresponding system of coupled Gross-Pitaevskii equations includes
fractional kinetic-energy operators, characterized by the L\'{e}vy index, $%
\alpha <2$ (the normal kinetic energy corresponds to $\alpha =2$). The SOC
terms, with strength $\lambda $, produce strong effects in the 2D case: they
create families of \emph{stable solitons} of the semi-vortex (SV) and\
mixed-mode (MM) types in the interval of $1<\alpha <2$, where the
supercritical collapse does not admit the existence of stable solitons in
the absence of the SOC. At $\lambda \rightarrow 0$, amplitudes of these
solitons vanish $\sim \lambda ^{1/(\alpha -1)}$.
\end{abstract}

\maketitle

\section{Introduction}

The concept of derivatives of fractional orders was proposed in 1823 by
Niels Henrik Abel \cite{Abel}. In physics, the implementation of the
fractionality was proposed by N. Laskin \cite{Laskin}, who had introduced a
linear fractional differential operator (actually, it is defined by an
integral expression, see Eq. (\ref{FracDefi}) below) which replaces the
usual quantum-mechanical kinetic energy. The result is the fractional Schr%
\"{o}dinger equation (FSE) for wave function $\psi (x,t)$, with the \textit{L%
\'{e}vy index} $\alpha $, time $t$, spatial coordinate $x$, and potential $%
U(x)$. The normalized form of the one-dimensional (1D) FSE is%
\begin{equation}
i\frac{\partial \psi }{\partial t}=\frac{1}{2}\left( -\frac{\partial ^{2}}{%
\partial x^{2}}\right) ^{\alpha /2}\psi +U(x)\psi .  \label{FSE}
\end{equation}%
It was derived, by means of the Feynman's path-integral method (tantamount
to stochastic quantization), for a particle whose stochastic motion is
realized not by the usual Brownian regime, but rather by \textit{L\'{e}vy
flights}. This means that the average distance of the corresponding randomly
walking classical particle from the initial position growth with time as $%
t^{1/\alpha }$ \cite{Levy flights}. In the case of $\alpha =2$, this is the
usual random-walk law for a Brownian particle, and, accordingly, Eq. (\ref%
{FSE}) amounts to the usual Schr\"{o}dinger equation. The L\'{e}vy-flight
regime, corresponding to $\alpha <2$, implies diffusive walk faster than
Brownian, which is performed, at the classical level, by random leaps. The
quantum counterpart of this regime, represented by Eq. (\ref{FSE}) in the 1D
case, is the basis of fractional quantum mechanics \cite{Laskin book}.

While there are different formal definitions of fractional derivatives, the
one which is relevant in the context of quantum mechanics is the \textit{%
Riesz derivative}, which is based on the juxtaposition of the direct and
inverse Fourier transforms \cite{Riesz-original,Riesz}:
\begin{equation}
\left( -\frac{\partial ^{2}}{\partial x^{2}}\right) ^{\alpha /2}\psi (x) =\frac{1%
}{2\pi }\int_{-\infty }^{+\infty }|p|^{\alpha }dp\int_{-\infty }^{+\infty
}\psi (\xi )\exp \left( ip(x-\xi )\right) d\xi .  \label{FracDefi}
\end{equation}%
Similarly, in the two-dimensional (2D) version of the FSE the fractional
kinetic-energy operator is defined as a fractional power of the Laplacian,
i.e.,%
\begin{gather}
\left( -\nabla ^{2}\right) ^{\alpha /2}\psi (x,y)\equiv \left( -\frac{\partial
^{2}}{\partial x^{2}}-\frac{\partial ^{2}}{\partial y^{2}}\right) ^{\alpha
/2}\psi  (x,y)\notag \\
=\frac{1}{(2\pi )^{2}}\int_{-\infty }^{+\infty }\int_{-\infty }^{+\infty
}\left( p^{2}+q^{2}\right) ^{\alpha /2}dpdq\int_{-\infty }^{+\infty
}\int_{-\infty }^{+\infty }\psi (\xi ,\eta )\exp \left[ip(x-\xi)
+iq(y-\eta) \right] d\xi d\eta ,  \label{2D operator}
\end{gather}

The similarity of the quantum-mechanical Schr\"{o}dinger equation to the
wave-propagation equation under the action of the paraxial diffraction in
optics suggests a possibility to simulate FSE in optical setups, with time $%
t $ replaced by the propagation distance, $z$. This option was proposed by
S. Longhi \cite{Longhi}, who elaborated a scheme based on a Fabry-Perot
resonator, in which the 4\textit{f} setup (two lenses with four focal
lengths \cite{4f}) is employed to approximate the fractional diffraction by
transforming the spatial structure of the light beam into its Fourier
counterpart, applying the fractional diffraction in its straightforward form
in the Fourier layer (as proposed in early works \cite{early1,early2}), and
then transforming the result back into the spatial domain. This physical
picture also corresponds to the definition of the Riesz derivative in Eqs. (%
\ref{FracDefi}) and (\ref{2D operator}). In other physical contexts, it was
proposed to create a \textit{L\'{e}vy crystal}, which provides
discretization of the FSE in a dynamical lattice \cite{Stickler}, and to
realize an effective FSE in a condensate of polaritons in a semiconductor
cavity \cite{Pinsker}.

The emulation of the fractional diffraction in optics suggests to take into
regard the natural Kerr nonlinearity of optical media, and thus to address a
possibility of the existence of fractional solitons. Various aspects of this
topic were a subject of many recent theoretical works \cite{Zhong}-\cite%
{Zeng 2}, see also a review in Ref. \cite{review}. In this case, the FSE in
free space (without the external potential) is replaced by one containing
the self-focusing cubic term:%
\begin{equation}
i\frac{\partial \psi }{\partial z}=\frac{1}{2}\left( -\nabla ^{2}\right)
^{\alpha /2}\psi -|\psi |^{2}\psi ,  \label{FNLSE}
\end{equation}%
where the fractional-diffraction operator $\left( -\nabla ^{2}\right)
^{\alpha /2}$ may be one- or two-dimensional, as defined in Eqs. (\ref%
{FracDefi}) and (\ref{2D operator}), respectively. In 1D, Eq. (\ref{FNLSE})
gives rise to stable solitons at $\alpha >1$ and to the \textit{critical
collapse} at $\alpha =1$, if the norm of a localized input exceeds a
critical value \cite{collapse,review}. In 2D, the critical collapse takes
place at $\alpha =2$ (it is the usual critical collapse in the 2D nonlinear
Schr\"{o}dinger equation with normal diffraction \cite{Sulem,Fibich}), and
the supercritical collapse (for which the critical value is zero) at $\alpha
<2$.

Another natural possibility\ is to consider a Bose-Einstein condensate (BEC)
in an ultracold gas of quantum particles moving by L\'{e}vy flights. In this
case, Eq. (\ref{FNLSE}), with $z$ replaced by time $t$, represents the
scaled form of the corresponding free-space Gross-Pitaevskii equation (GPE),
assuming attractive interaction between particles. As well as in the
realization in optics, a crucially important issue is a possibility of the
stabilization of solutions for matter-wave solitons, created by Eq. (\ref%
{FNLSE}), against the collapse. In this connection, it is relevant to
mention that the spin-orbit coupling (SOC) efficiently stabilizes 2D
matter-wave solitons of two types, \textit{viz}., semi-vortices (SVs) and
mixed modes (MMs), against the critical collapse, in the framework of the
two-component system of GPEs with the normal kinetic-energy operators and
cubic self- and cross-attraction terms \cite{we,Sherman}, see also a review
in Ref. \cite{EPL}. SVs are composite states with zero vorticity in one
component, and vorticity $1$ in the other, while MMs mix terms with zero and
nonzero vorticities in both components. Thus, a relevant issue is to
consider the SOC system with the fractional kinetic-energy operators, and
construct soliton states in them, with emphasis on possibilities to
stabilize them against the collapse. This is the subject of the present work.

The character of the motion of the particles in the condensate (L\'{e}vy
flights) does not affect the form of the nonlinearity in the GPE. As
concerns the SOC\ term in the system's Hamiltonian, its systematic
derivation, starting from the consideration of transitions between two
intrinsic quantum states in the particle, driven by the Raman laser fields,
and applying the unitary transformation to the spinor wave function \cite%
{Zhai,Busch}, produces a cumbersome result, in the case of the fractional
kinetic energy. In this work, we adopt a phenomenological model, in which
the SOC is represented by the same term of the Rashba type in the system's
Hamiltonian as in the usual situation \cite{usual}. Actually, the results
reported below demonstrate that the SOC produces a minor effect in the 1D
setup (therefore, the 1D case is considered below in a brief form in Section
2), while in the most interesting 2D case, considered in Section 3, the
effect is conspicuous: it shows partial stabilization of the solitons of the
SV\ and MM types at $1<\alpha \leq 2$ (the range of $\alpha <1$ is
considered too, but it produces no soliton states).

It is plausible that the results will be quite similar beyond the framework
of the phenomenological model. Indeed, the stabilization of the 2D solitons
by SOC is accounted for by its spatial scale. It breaks the scaling
invariance which underlies the onset of the collapse \cite{Sulem,Fibich}.
This fact is equally true for the full and phenomenological forms of the
system which includes the fractional kinetic energy and SOC interaction. On
the other hand, it is expected that the cumbersome form of the full system
will not allow one to clearly distinguish the SV and MM species of 2D
solitons, adding extra vortical terms to both components of each soliton.

In the analysis following below, stationary shapes of both 1D and 2D
solitons are produced in a numerical form, and, in parallel, analytically by
means of the variational approximation (VA). Stability of the solitons is
tested with the help of direct simulations of their perturbed evolution.

\section{The 1D system}

The system of 1D GPEs for the spinor wave function, $\boldsymbol{\phi }%
=(\phi _{+},\phi _{-})$, of the binary BEC with attractive contact
interactions and the phenomenologically introduced SOC terms of the Rashba
type \cite{usual}, with strength $\lambda $, under the action of the
fractional kinetic-energy operator (\ref{FracDefi}) is written, in the
scaled form, as
\begin{eqnarray}
i\frac{\partial \phi _{+}}{\partial t} &=&\frac{1}{2}\left( -\frac{\partial
^{2}}{\partial x^{2}}\right) ^{\alpha /2}\phi _{+}-(|\phi _{+}|^{2}+\gamma
|\phi _{-}|^{2})\phi _{+}+\lambda \frac{\partial \phi _{-}}{\partial x},
\notag \\
i\frac{\partial \phi _{-}}{\partial t} &=&\frac{1}{2}\left( -\frac{\partial
^{2}}{\partial x^{2}}\right) ^{\alpha /2}\phi _{-}-(|\phi _{-}|^{2}+\gamma
|\phi _{+}|^{2})\phi _{-}-\lambda \frac{\partial \phi _{+}}{\partial x},
\label{1D}
\end{eqnarray}%
where $\gamma \geq 0$ is the relative strength of the inter-component
attraction, while the strength of the self-attraction is scaled to be $1$.
Stationary soliton solutions of Eq. (\ref{1D}) with chemical potential $\mu
<0$ are looked for as
\begin{equation}
\phi _{\pm }\left( x,t\right) =e^{-i\mu t}u_{\pm }(x),  \label{mu}
\end{equation}%
with real functions $u_{\pm }(x)$ satisfying the following equations:%
\begin{eqnarray}
\mu u_{+} &=&\frac{1}{2}\left( -\frac{\partial ^{2}}{\partial x^{2}}\right)
^{\alpha /2}u_{+}-(u_{+}^{2}+\gamma u_{-}^{2})u_{+}+\lambda \frac{du_{-}}{dx}%
,  \notag \\
\mu u_{-} &=&\frac{1}{2}\left( -\frac{\partial ^{2}}{\partial x^{2}}\right)
^{\alpha /2}u_{-}-(u_{-}^{2}+\gamma u_{+}^{2})u_{-}-\lambda \frac{du_{+}}{dx}%
.  \label{u}
\end{eqnarray}%
In agreement Eqs. (\ref{u}), it is possible to set $u_{+}(x)$ and $u_{-}(x)$
to be, respectively, even and odd functions of $x$.

Note that the system of 1D stationary equations (\ref{u}) can be derived
from the Lagrangian,%
\begin{gather}
L_{\mathrm{1D}}=\frac{\mu }{2}\int_{-\infty }^{+\infty }dx\left[
u_{+}^{2}(x)+u_{-}^{2}(x)\right]   \notag \\
-\frac{1}{4\pi }\int_{0}^{+\infty }p^{\alpha }dp\int_{-\infty }^{+\infty
}d\xi \int_{-\infty }^{+\infty }dx\cos \left( p(x-\xi \right) )\cdot \left[
u_{+}(x)u_{+}(\xi )+u_{-}(x)u_{-}(\xi )\right]   \notag \\
+\int_{-\infty }^{+\infty }\left[ \lambda u_{-}\frac{du_{+}}{dx}+\frac{1}{4}%
\left( u_{+}^{4}+u_{-}^{4}\right) +\frac{\gamma }{2}u_{+}^{2}u_{-}^{2}\right]
dx.  \label{L}
\end{gather}%
Taking into regard the parities of functions $u_{\pm }(x)$, the Lagrangian
can be additionally simplified:%
\begin{gather}
L_{\mathrm{1D}}=\frac{\mu }{2}\int_{-\infty }^{+\infty }dx\left[
u_{+}^{2}(x)+u_{-}^{2}(x)\right]   \notag \\
-\frac{1}{4\pi }\int_{0}^{+\infty }p^{\alpha }dp\int_{-\infty }^{+\infty
}d\xi \int_{-\infty }^{+\infty }dx\left\{ \cos (px)\cos (p\xi )\left[
u_{+}(x)u_{+}(\xi )\right] +\sin (px)\sin (p\xi )\left[ u_{-}(x)u_{-}(\xi )%
\right] \right\}   \notag \\
+\int_{-\infty }^{+\infty }dx\left[ \lambda u_{-}\frac{du_{+}}{dx}+\frac{1}{4%
}\left( u_{+}^{4}+u_{-}^{4}\right) +\frac{\gamma }{2}u_{+}^{2}u_{-}^{2}%
\right] .  \label{Lagr}
\end{gather}

The Lagrangian is used to look for solitons in the framework of the VA,
which may be based on the Gaussian-shaped ansatz,
\begin{equation}
u_{+}(x)=\sqrt{\frac{N}{\sqrt{\pi }W}}\left( \cos \theta \right) \exp \left(
-\frac{x^{2}}{2W^{2}}\right) ,u_{-}(x)=\sqrt{\frac{2N}{\sqrt{\pi }W}}\left(
\sin \theta \right) \frac{x}{W}\exp \left( -\frac{x^{2}}{2W^{2}}\right) .
\label{ansatz}
\end{equation}%
Free parameters $\theta $ and $W$ in Eq. (\ref{ansatz}) determine the
distribution of the norm between the components and the width of the
soliton, for fixed total norm,
\begin{equation}
N=\int_{-\infty }^{+\infty }dx\left( u_{+}^{2}+u_{-}^{2}\right) .  \label{N}
\end{equation}%
The substitution of ansatz (\ref{ansatz}) in Lagrangian (\ref{Lagr}) yields
an expressions for the effective Lagrangian,%
\begin{eqnarray}
\left( L_{\mathrm{eff}}\right) _{\mathrm{1D}} &=&\frac{\mu }{2}N-\Gamma
\left( \frac{\alpha +1}{2}\right) \frac{N}{4\sqrt{\pi }W^{\alpha }}\left[
\cos ^{2}\theta +\left( \alpha +1\right) \sin ^{2}\theta \right] -\frac{%
\lambda N}{2\sqrt{2}W}\sin \left( 2\theta \right)   \notag \\
&&+\frac{N^{2}}{4\sqrt{2\pi }W}\left( \cos ^{4}\theta +\frac{3}{4}\sin
^{4}\theta \right) +\frac{\gamma N^{2}}{16\sqrt{2\pi }W}\sin ^{2}(2\theta ),
\label{Leff}
\end{eqnarray}%
where $\Gamma $ is the Euler's Gamma-function. Then, for given $N$ the VA
predicts values $\theta $ and $W$ as solutions of the Euler-Lagrange
equations,
\begin{equation}
\frac{\partial \left( L_{\mathrm{eff}}\right) _{\mathrm{1D}}}{\partial W}=%
\frac{\partial \left( L_{\mathrm{eff}}\right) _{\mathrm{1D}}}{\partial
\theta }=0.  \label{EL}
\end{equation}%
In the case of $\lambda =0$ in Eq. (\ref{1D}) and, accordingly, $\theta =0$
in Eq. (\ref{ansatz}), equation $\partial \left( L_{\mathrm{eff}}\right) _{%
\mathrm{1D}}/\partial W=0$ is identical to that derived in Ref. \cite%
{Guangzhou}. In addition to Eq. (\ref{EL}), a relation between $\mu $ and $N$
is determined, in the framework of the VA, by equation%
\begin{equation}
\frac{\partial \left( L_{\mathrm{eff}}\right) _{\mathrm{1D}}}{\partial N}=0.
\label{EL-N}
\end{equation}

Note that the equations which correspond to the critical collapse, i.e., the
1D version of Eq. (\ref{FNLSE}) with $\alpha =1$, or its 2D version with $%
\alpha =2$, produce degenerate families of (unstable) soliton solutions,
with the single (critical) value of the norm, $N_{c}$, that does not depend
on $\mu $. In particular, in the 2D case with $\alpha =2$, this is the
well-known family of \textit{Townes solitons} \cite{Townes,Sulem,Fibich},
whose single (numerically found) value of the norm is
\begin{equation}
\left( N_{c}\right) _{\mathrm{2D}}\equiv \left( N_{\mathrm{Townes}}^{\mathrm{%
(num)}}\right) _{\mathrm{2D}}\approx 5.85.  \label{Townes-num}
\end{equation}%
The simple Gaussian-based VA for the Townes solitons predicts an approximate
value of the norm \cite{Anderson},%
\begin{equation}
\left( N_{\mathrm{Townes}}^{\mathrm{(VA)}}\right) _{\mathrm{2D}}=2\pi ,
\label{Townes-VA}
\end{equation}%
the relative error, in comparison to Eq. (\ref{Townes-num}) being $\simeq 7\%
$. In the case of the 1D version of Eq. (\ref{FNLSE}) with $\alpha =1$,
similar results for the degenerate soliton family are \cite{Zeng 2}
\begin{equation}
~\left( N_{c}\right) _{\mathrm{1D}}\equiv N_{\mathrm{1D,}\alpha =1}^{\mathrm{%
(num)}}\approx 1.23,~N_{\mathrm{1D,}\alpha =1}^{\mathrm{(VA)}}=\sqrt{2},
\label{Townes}
\end{equation}%
the relative error being $\simeq 13\%$, cf. Eqs. (\ref{Townes-num}) and (\ref%
{Townes-VA}).

If the SOC terms are included ($\lambda >0$), width $W$ drops out from Eqs. (%
\ref{EL}) and (\ref{Leff}) at $\alpha =1$, hence this system too admits a
solution at a single (\textit{critical}) value $N_{c}$ of $N$, i.e., the
soliton family remains a degenerate one in this case. This conclusion, which
is a natural consequence of the fact that operator (\ref{FracDefi}) with $%
\alpha =1$ and SOC terms in Eq. (\ref{1D}) feature the same scaling with
respect to $x$, is confirmed by a numerically found solution of Eq. (\ref{u}%
), which was obtained by applying the imaginary-time integration method to
the full system of equations (\ref{1D}).

\begin{figure}[h]
\begin{center}
\includegraphics[height=4.cm]{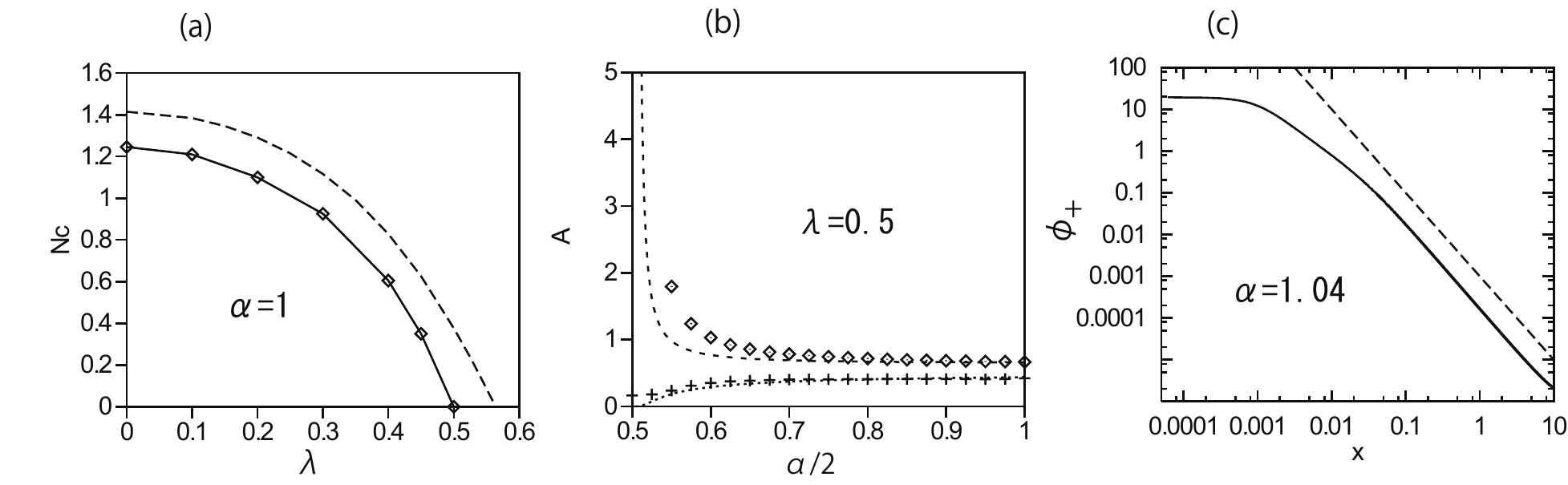}
\end{center}
\caption{(a) The chain of rhombuses shows the numerically found critical
norm $N_{c}$ of the degenerate 1D soliton family as a function of the SOC
strength, $\protect\lambda $, at $\protect\alpha =1$. The numerical results
are produced by imaginary-time simulations of Eq. (\protect\ref{1D}). The
dashed line is the same value as predicted by the VA, as per Eqs. (\protect
\ref{EL}) and (\protect\ref{Leff}). The numerical and variational values of $%
N_{c}$ vanish precisely at points (\protect\ref{max_exact}) and (\protect\ref%
{lambda_max}), respectively. (b) Chains of rhombuses and crosses show the
amplitude of component $|\protect\phi _{+}(x)|$ of the 1D solitons for two
fixed values of the norm, $N=1.3$ and $0.8$, respectively, as functions of $%
\protect\alpha /2$ at fixed $\protect\lambda =0.3$. Dashed lines represent
the respective VA-predicted results. (c) An example of a stable 1D soliton,
found in the numerical form at $\protect\alpha =1.04$ and $\protect\lambda %
=1/2$, with total norm $N=1$, is displayed by means of the
double-logarithmic plot of $|\protect\phi _{+}(x)|$. The dashed-line fit is $%
0.001/x^{2}$ (this fit is not valid at still larger values of $|x|$, where
the soliton's tail decays exponentially). In this figure and Fig. \protect
\ref{fig2}, the results are presented for $\protect\gamma =0$ in Eq. (%
\protect\ref{1D}) (no attraction between the two components).}
\label{fig1}
\end{figure}

Dependences $N_{c}(\lambda )$ for the 1D system with $\alpha =1$, as
produced by the numerical solution and VA, are displayed in Fig. \ref{fig1}%
(a). In particular, a straightforward consequence of Eqs. (\ref{EL}) and (%
\ref{Leff}) is the value of the SOC strength, $\lambda $, at which the
VA-predicted critical norm vanishes:%
\begin{equation}
N_{c}^{\mathrm{(var)}}\left( \lambda =\left( \lambda _{\max }^{\mathrm{(var)}%
}\right) _{\alpha =1}\equiv 1/\sqrt{\pi }\approx 0.56\right) =0,
\label{lambda_max}
\end{equation}%
with no solitons existing at $\lambda >\lambda _{\max }$. The exact value of
$\lambda _{\max }$ can be found too, without the use of the VA. Indeed, it
follows from the consideration of the scaling of soliton's tails decaying at
$|x|\rightarrow \infty $, which are produced by Eq. (\ref{u}) with $\alpha =1
$, that exponentially decaying tails are possible at%
\begin{equation}
\lambda <\left( \lambda _{\max }^{\mathrm{(exact)}}\right) _{\alpha =1}=1/2.
\label{max_exact}
\end{equation}%
At $\lambda >1/2$, there exist solutions for tails decaying algebraically,
rather than exponentially:
\begin{equation}
u_{+,-}(x)\sim |x|^{-1/\left( 4\lambda ^{2}-1\right) }.  \label{tails}
\end{equation}%
The results demonstrate that slowly decaying tails (\ref{tails}) cannot be
used for constructing solitons (in addition, it is relevant to mention that
the total norm, determined by the asymptotic form (\ref{tails}) at $%
|x|\rightarrow \infty $, diverges at $\lambda \geq \sqrt{3}/2$).

The norm-degenerate family of the 1D solitons existing at $\alpha =1$ and $%
\lambda <1/2$ is completely unstable. Its relation to nondegenerate families
of stable solitons existing at $\alpha >1$ is illustrated by Fig. \ref{fig1}%
(b), which shows the amplitude of the solitons as a function of $\alpha $
for a fixed SOC strength, $\lambda =0.3$, which is smaller than the one
given by Eq. (\ref{max_exact}) (at $\lambda >1/2$, the solitons definitely
do not exist in the limit of $\alpha =1$, as shown above), and for two
different fixed norms, $N=0.8$ and $1.3$. Because both are different from
the respective value $N_{c}^{\mathrm{(num)}}(\lambda =0.3)\approx 0.9$, see
Fig. \ref{fig1}(a), the 1D solitons with these fixed values of $N$ do not
exist in the limit of $\alpha =1$. Accordingly, Fig. \ref{fig1}(b)
demonstrates that the solitons with $N<N_{c}^{\mathrm{(num)}}(\lambda =0.3)$
suffer delocalization at $\alpha \rightarrow 1$, with the amplitude
approaching zero in this limit, while the solitons with $N>N_{c}^{\mathrm{%
(num)}}(\lambda =0.3)$ suffer collapse (divergence of the amplitude).

\begin{figure}[h]
\begin{center}
\includegraphics[height=4.cm]{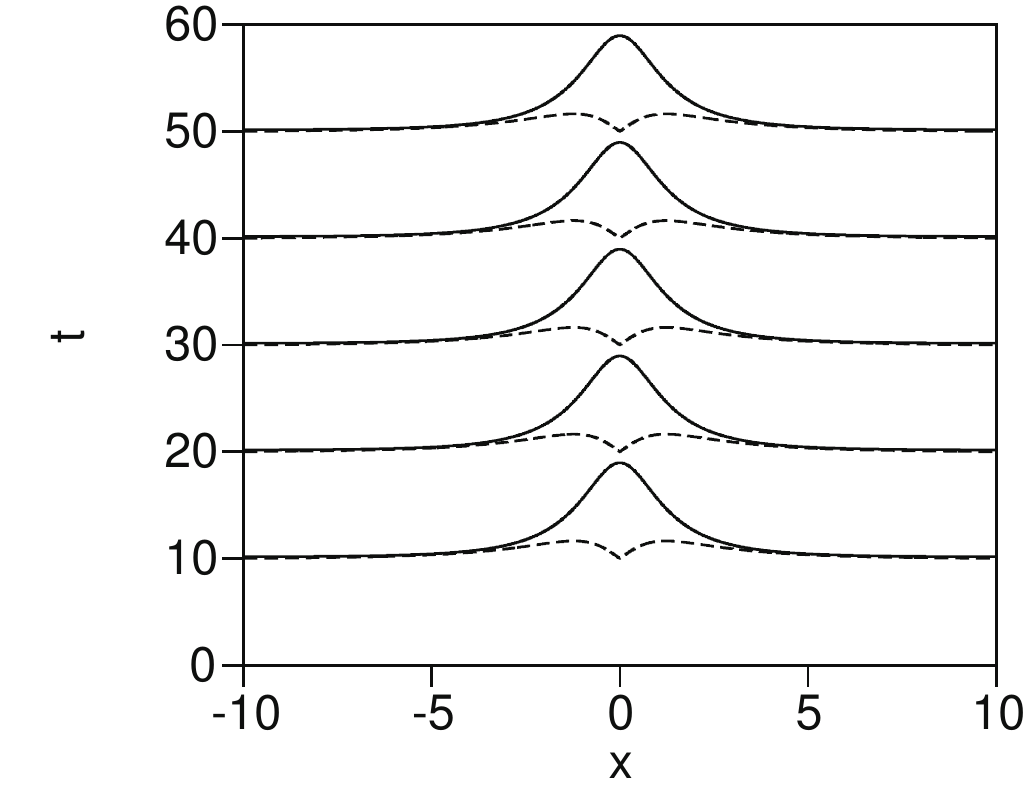}
\end{center}
\caption{The evolution of $|\protect\phi _{+}(x)|$ and $|\protect\phi %
_{-}(x)|$ (solid and dashed lines, respectively), produced by simulations of
Eq. (\protect\ref{1D}) with $\protect\alpha =1.4$ and $\protect\lambda =0.3$%
. The norm of this 1D soliton is $N=1.3$.}
\label{fig2}
\end{figure}

Figure \ref{fig1}(c) shows that, at values of $\alpha $ slightly exceeding $1
$ (here, $\alpha =1.04$) and $\lambda =1/2$, the 1D soliton exists with tails close to the algebraic shape (comparison to that expression is relevant as $\alpha$ is close to $1$, although Eq.~(20) cannot be directly applied because $\alpha$ is not 1). Lastly, Fig. \ref{fig2} illustrates
the stability of a typical 1D soliton with $N=1.3$ at $\alpha =1.4$ and $%
\lambda =0.3$.

Thus, the effect of the SOC terms in the 1D system changes the critical norm for the collapse but fails to stabilize the solitons in the critical case of $\alpha =1$. At $\alpha >1$, the effect is not strong either. In particular, additional calculations
clearly demonstrate that variation of $\alpha $ does not strongly affect
stability of the 1D solitons (details are not displayed here, as they do not
convey something remarkable). It is shown below that the SOC terms produce a
much more conspicuous impact in the 2D system.

\section{The 2D system}

Following Ref. \cite{we}, the 2D system of GPEs with the SOC coupling of the
Rashba type and the fractional kinetic-energy operator (\ref{2D operator})
is written, in the scaled form, as
\begin{eqnarray}
i\frac{\partial \phi _{+}}{\partial t} &=&\frac{1}{2}\left( -\nabla
^{2}\right) ^{\alpha /2}\phi _{+}-(|\phi _{+}|^{2}+\gamma |\phi
_{-}|^{2})\phi _{+}+\lambda \left( \frac{\partial \phi _{-}}{\partial x}-i%
\frac{\partial \phi _{-}}{\partial y}\right) ,  \notag \\
i\frac{\partial \phi _{-}}{\partial t} &=&\frac{1}{2}\left( -\nabla
^{2}\right) ^{\alpha /2}\phi _{-}-(|\phi _{-}|^{2}+\gamma |\phi
_{+}|^{2})\phi _{-}-\lambda \left( \frac{\partial \phi _{+}}{\partial x}+i%
\frac{\partial \phi _{+}}{\partial y}\right) ,  \label{2D}
\end{eqnarray}%
cf. Eq. (\ref{1D}). Note that, except for case of $\alpha =1$, rescaling
\begin{equation}
\left( x,y\right) =\lambda ^{-1/(\alpha -1)}\left( \tilde{x},\tilde{y}%
\right) ,t=\lambda ^{-\alpha /(\alpha -1)}\tilde{t},\phi _{\pm }=\lambda
^{\alpha /\left( 2(\alpha -1)\right) }\tilde{\phi}_{\pm }  \label{rescaling}
\end{equation}%
makes it possible to set $\lambda \equiv 1$ in Eq. (\ref{2D}). Accordingly,
the norm of the 2D states scales with the variation of the SOC strength as%
\begin{equation}
N(\lambda )\equiv \int \int \left( \left\vert u_{+}\right\vert
^{2}+\left\vert u_{-}\right\vert ^{2}\right) dxdy=\lambda ^{-(2-\alpha
)/(\alpha -1)}N(\lambda =1),  \label{N(lambda)}
\end{equation}%
where $N(\lambda= 1)$ is the norm at the reference point, $\lambda=1$.

Solutions of Eqs. (\ref{2D}) for 2D stationary states with chemical
potential $\mu $ are looked for as
\begin{equation}
\phi _{\pm }\left( x,y,t\right) =e^{-i\mu t}u_{\pm }(x,y),  \label{mu2D}
\end{equation}%
with complex functions $u_{\pm }(x,y)$, cf. the 1D counterpart given by Eq. (%
\ref{mu}) with real functions $u_{\pm }(x)$. The substitution of ansatz (\ref%
{mu2D}) in Eqs. (\ref{2D}) leads to the stationary equations for complex
functions $u_{\pm }(x,y)$:%
\begin{eqnarray}
\mu u_{+} &=&\frac{1}{2}\left( -\nabla ^{2}\right) ^{\alpha
/2}u_{+}-(|u_{+}|^{2}+\gamma |u_{-}|^{2})u_{+}+\lambda \left( \frac{\partial
u_{-}}{\partial x}-i\frac{\partial u_{-}}{\partial y}\right) ,  \notag \\
\mu u_{-} &=&\frac{1}{2}\left( -\nabla ^{2}\right) ^{\alpha
/2}u_{-}-(|u_{-}|^{2}+\gamma |u_{+}|^{2})u_{-}-\lambda \left( \frac{\partial
u_{+}}{\partial x}+i\frac{\partial u_{+}}{\partial y}\right) ,  \label{u2D}
\end{eqnarray}%
cf. the 1D stationary equations (\ref{u}). Taking into regard definition (%
\ref{2D operator}), Eqs. (\ref{u2D}) can be derived from the respective
Lagrangian:%
\begin{gather}
L=\mu \int_{-\infty }^{+\infty }dx\int_{-\infty }^{+\infty }dy\left[
\left\vert u_{+}(x,y)\right\vert ^{2}+\left\vert u_{-}(x,y)\right\vert ^{2}%
\right]   \notag \\
-\frac{1}{2\pi ^{2}}\int_{0}^{+\infty }dp\int_{0}^{+\infty
}dq(p^{2}+q^{2})^{\alpha /2}\int_{-\infty }^{+\infty }d\xi \int_{-\infty
}^{+\infty }d\eta \int_{-\infty }^{+\infty }dx\int_{-\infty }^{+\infty
}dy\cos \left[ p(x-\xi )+q(y-\eta )\right]   \notag \\
\times \left[ u_{+}^{\ast }(x,y)u_{+}(\xi ,\eta )+u_{-}^{\ast
}(x,y)u_{-}(\xi ,\eta )\right]   \notag \\
-\lambda \int_{-\infty }^{+\infty }dx\int_{-\infty }^{+\infty }dy\left[
u_{+}^{\ast }\frac{\partial u_{-}}{\partial x}+u_{+}\frac{\partial
u_{-}^{\ast }}{\partial x}-i\left( u_{+}^{\ast }\frac{\partial u_{-}}{%
\partial y}-u_{+}\frac{\partial u_{-}^{\ast }}{\partial y}\right) \right]
\notag \\
+\int_{-\infty }^{+\infty }dx\int_{-\infty }^{+\infty }dy\left[ \frac{1}{2}%
\left( |u_{+}|^{4}+|u_{-}|^{4}\right) +\gamma |u_{+}|^{2}|u_{-}|^{2}\right] ,
\label{L2D}
\end{gather}%
cf. Eq. (\ref{L}). The integration with respect to $p$ and $q$ in Eq. (\ref%
{L2D}) is reduced from original domains $\left( -\infty ,+\infty \right) $
to $\left( 0,\infty \right) $, using the symmetry of the domain.

As suggested by Ref. \cite{we} (which addressed the system with the normal
kinetic-energy operator, i.e., $\alpha =2$), we first aim to construct 2D
soliton solutions of the SV type. To this end, the following ansatz is
adopted:
\begin{equation}
u_{+}=A_{+}\exp (-\beta (x^{2}+y^{2})),u_{-}=A_{-}(x+iy)\exp (-\beta
(x^{2}+y^{2})),  \label{ans2D}
\end{equation}%
which implies vorticities $0$ and $1$ in the components $u_{+}$ and $u_{-}$,
respectively, in accordance with the definition of the SV mode. Note that
the maximum of component $u_{-}$ of the ansatz, $\left\vert
u_{-}(r)\right\vert _{\max }=e^{-1/2}A_{-}/\sqrt{2\beta }$, is attained at $%
r=1/\sqrt{2\beta }$.

The substitution of ansatz (\ref{ans2D}) in Lagrangian (\ref{L2D}) produces
the respective effective Lagrangian, cf. Eq. (\ref{Leff}):
\begin{equation}
L_{\mathrm{eff}}=\mu N-\frac{\pi \Gamma \left( 1+\alpha /2\right) }{2\left(
2\beta \right) ^{1-\alpha /2}}\left[ A_{+}^{2}+\left( 1+\frac{\alpha }{2}%
\right) \frac{A_{-}^{2}}{2\beta }\right] -\frac{\pi \lambda }{\beta }%
A_{+}A_{-}+\frac{\pi }{8\beta }\left( A_{+}^{4}+\frac{A_{-}^{4}}{8\beta ^{2}}%
+\frac{\gamma A_{+}^{2}A_{-}^{2}}{2\beta }\right) ,  \label{va1}
\end{equation}%
where the total 2D norm of ansatz (\ref{ans2D}) is
\begin{equation}
N=\frac{\pi }{2\beta }\left( A_{+}^{2}+\frac{A_{-}^{2}}{2\beta }\right) .
\label{N2D}
\end{equation}%
Then, for given $\mu $, SV's\ parameters are predicted by the Euler-Lagrange
equations (cf. Eq. (\ref{EL})),%
\begin{equation}
\frac{\partial L_{\mathrm{eff}}}{\partial A_{\pm }}=\frac{\partial L_{%
\mathrm{eff}}}{\partial \beta }=0.  \label{EL2D}
\end{equation}

\begin{figure}[h]
\begin{center}
\includegraphics[height=5.cm]{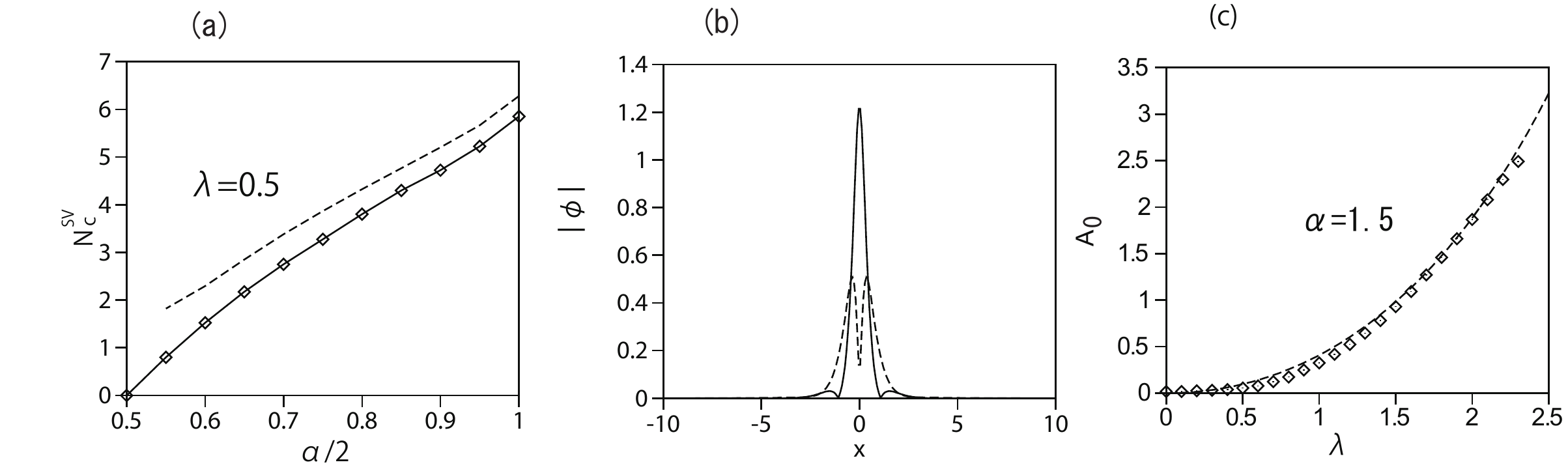}
\end{center}
\caption{(a) The critical value of the norm of the SV solitons in the 2D
system, $N_{c}^{\mathrm{(SV)}}$, as a function of the half L\'{e}vy index, $%
\protect\alpha /2$, at $\protect\lambda =0.5$ for $\protect\gamma =0$ in Eq.
(\protect\ref{2D}). At $1<\protect\alpha \leq 2$, the system gives rise to
stable SV (semi-vortex) solitons at $N<N_{c}^{\mathrm{(SV)}}$, and to
collapse at $N>N_{c}^{\mathrm{(SV)}}$. Here and in panel (c), the chain of
rhombuses and dashed line represent, respectively, the numerical findings,
obtained from the imaginary-time simulations of Eq. (\protect\ref{2D}), and
VA results produced by Eqs.~(\protect\ref{va1}) and (\protect\ref{EL2D}).
(b) The cross-section profiles of $|\protect\phi _{+}|$ and $\left\vert
\protect\phi _{-}\right\vert $ (the solid and dashed lines, respectively),
produced at $\protect\lambda =0.5$ and $\protect\alpha =1$ by the numerical
solution of the \emph{linearized system} of size $20\times 20$. Amplitudes
of the linear solution are determined by an (arbitrarily adopted)
normalization condition $N=1$. (c) The amplitude of the SV solitons (largest
value of $|\protect\phi _{+}(x)|$ vs. the SOC strength $\protect\lambda $,
at $\protect\alpha =1.5$, $\protect\gamma =0$ and a fixed total norm, $N=0.5$%
. At small $\protect\lambda $, the dependence is explained by the scaling
relation (\protect\ref{scaling}) with $\protect\alpha =1.5$, i.e., $%
A_{0}\sim \protect\sqrt{N}\protect\lambda ^{2}$. }
\label{fig3}
\end{figure}

Figure \ref{fig3} presents the most essential results of the analysis for
the 2D case with $\gamma =0$ (no nonlinear interaction between components $%
\phi _{\pm }$ in Eq. (\ref{2D})). Namely, unlike the 2D version of Eq. (\ref%
{FNLSE}) with $\alpha \leq 2$, which does not include SOC and generates only
completely unstable soliton solutions, Eq. (\ref{2D}) gives rise to \emph{%
stable SVs} at $1<\alpha \leq 2$, with norms falling below the respective
critical value, \textit{viz}., $N\,<N_{c}^{\mathrm{(SV)}}(\alpha )$. The
dependence of $N_{c}^{\mathrm{(SV)}}$ on $\alpha $, as obtained from the
numerical solution (using simulations of Eq. (\ref{2D}) in imaginary time)
and VA, is displayed in Fig. \ref{fig3}(a) for $\lambda =0.5$. As it follows
from the above argument which demonstrates the peculiarity of the case of $%
\alpha =1$, $N_{c}^{\mathrm{(SV)}}$ vanishes at $\alpha =1$, which is indeed
demonstrated by the numerical results in Fig. \ref{fig3}(a) (while the VA
result is inaccurate at $\alpha \rightarrow 1$, as the ansatz (\ref{ans2D})
is not relevant in this limit). Thus, SV states with a finite amplitude do
not exist at $\alpha =1$. On the other hand, for the same case of $\alpha =1$
the solution with the vanishing amplitude can be obtained from the
linearized version of Eq. (\ref{2D}). This solution, generated by the
imaginary-time simulations of the linearized equations, with periodic
boundary conditions fixed for domain $|x|,|y|\leq 10$, is displayed by means
of cross sections $\left\vert \phi _{\pm }\left( x\right) \right\vert $,
drawn through $y=0$, in Fig. \ref{fig3}(b). The effective localization of
the solution implies that it is, roughly speaking, similar to periodic
states represented by the Jacobi's elliptic functions with modulus close to $%
k=1$.

The build of the family of stable SV solitons, constructed by means of the
numerical solution and VA, is illustrated by Fig. \ref{fig3}(c) for a
generic value of the L\'{e}vy index, $\alpha =1.5$. In this figure, the SOC
strength $\lambda $ is not fixed but varied, to explicitly display the
effect of the SOC, while the scaling is used to keep a fixed value of the
norm, $N=0.5$ (the mutual scaling of $N$ and $\lambda $ is determined by Eq.
(\ref{N(lambda)}). It is seen that the SV's amplitude, $A_{0}(\lambda )$,
vanishes at $\lambda \rightarrow 0$ (in agreement with the fact there are no
stable 2D solitons in the absence of the SOC in the system). Straightforward
analysis of Eqs. (\ref{u2D}) and its VA counterpart (\ref{EL2D})
demonstrates the following scaling relation between the SV's amplitude, $%
A_{0}$, its width, $\beta ^{-1/2}$ (see Eq. (\ref{ans2D})), and $\lambda $,
at $\lambda \rightarrow 0$ (cf. Eq. (\ref{N(lambda)}):
\begin{equation}
A_{0}\sim \sqrt{N}\lambda ^{1/(\alpha -1)},\beta ^{-1/2}\sim \lambda
^{-1/(\alpha -1)}.  \label{scaling}
\end{equation}%
For $\alpha =1.5$, Eq. (\ref{scaling}) yields $A_{0}\sim \sqrt{N}\lambda ^{2}
$, which obviously agrees with the curves plotted in Fig. \ref{fig3}(c).

Further, Fig. \ref{fig3}(c) shows that the amplitude increases with the
increase of the SOC strength, up to a maximum value, $\lambda _{\max
}\approx 2.4$, above which the SV soliton is destroyed by the collapse.
While the VA produces, in the same figure, the amplitude-vs.-$\lambda $
curve which is very close to the numerical counterpart, it does not
terminate at $\lambda $ close to $\lambda _{\max }$. Nevertheless, the
termination of the soliton family at large $\lambda $ is a natural finding,
as the existence of solitons in the system with the SOC terms dominating
over the kinetic energy requires the presence of the Zeeman splitting
between the components \cite{we2}, which is absent in Eqs. (\ref{1D}) and (%
\ref{2D}).

It is plausible that the stable SV solitons, populating the area beneath the
right branch of the $N_{c}^{\mathrm{(SV)}}(\alpha )$ curve in Fig. \ref{fig3}%
(a), play the role of the system's ground state, which is missing (formally
replaced by the collapsing solution) in the absence of the SOC terms, cf. a
similar conclusion made in Ref. \cite{we} for the usual 2D system with $%
\alpha =2$. This point calls for a deeper consideration, which is beyond the
scope of the present paper.

At values of the norm $N>N_{c}^{\mathrm{(SV)}}$ simulations of Eqs. (\ref{2D}%
) demonstrate the collapse of the SV, no stationary solitons being possible.
At $\alpha <1$, the system does not give rise to solitons either.

\begin{figure}[h]
\begin{center}
\includegraphics[height=4.cm]{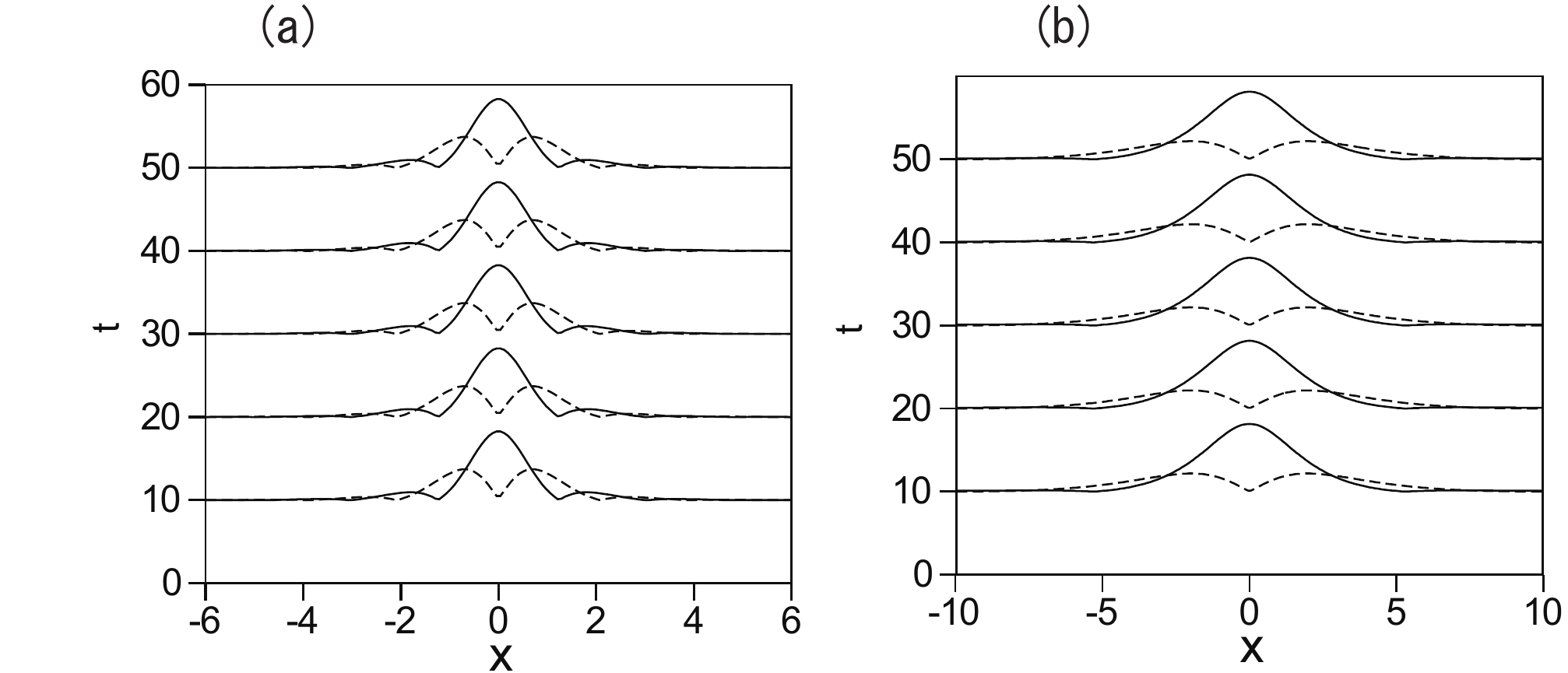}
\end{center}
\caption{Stable evolution of $|\protect\phi _{+}(x,y)|$ and $|\protect\phi %
_{-}(x,y)|$ in the SV soliton, plotted in the cross section $y=0$ by solid
and dashed lines, respectively, (a) An example for a typical fractional
system, with $\protect\lambda =1$, $\protect\alpha =1.5$, and $\protect%
\gamma =0$ in Eq. (\protect\ref{2D}), and the soliton's norm $N=1$. (b) The
stable evolution in a system close to the regular (non-fractional) one, with
$\protect\lambda =0.4$, $\protect\alpha =1.9$, $\protect\gamma =0$, and $%
N=5.15$.}
\label{fig5}
\end{figure}

The evolution and stability of typical SV solitons in real-time simulations
of Eq. (\ref{2D}) with $\gamma =0$ are illustrated in Fig. \ref{fig5} for
(a) $\lambda =1$, $\alpha =1.5$, and $N=1$ and (b) $\lambda =0.4$, $\alpha
=1.9$, and $N=5.15$. The latter case represents the 2D system with the L\'{e}%
vy index close to $\alpha =2,$ which corresponds to the system with the
usual (non-fractional) kinetic-energy operator, that was considered in Ref.
\cite{we}.

The above consideration was performed for $\gamma =0$ in Eq. (\ref{2D}),
when (as well as for $\gamma <1$, i.e., for the self-attraction of each
component stronger than the cross-attraction) the SV is expected to be the
dominant species of 2D solitons \cite{we}. On the other hand, in the usual
(non-fractional) system, SVs are unstable at $\gamma >1$, while stable 2D
solitons are MM solution, which, as mentioned above, mix the zero-vorticity
and vortex terms in each component. The MM represents the ground state of
the non-fractional system with $\gamma >1$ \cite{we}.

In the present case, the MMs may be approximated by the following
variational ansatz for the stationary components (see Eq. (\ref{mu2D})),
\begin{eqnarray}
u_{+} &=&A_{1}\exp (-\beta (x^{2}+y^{2}))-A_{2}(x-iy)\exp (-\beta
(x^{2}+y^{2})),  \notag \\
u_{-} &=&A_{1}\exp (-\beta (x^{2}+y^{2}))+A_{2}(x+iy)\exp (-\beta
(x^{2}+y^{2})),  \label{MM}
\end{eqnarray}%
which indeed mixes terms with zero and nonzero vorticities. The substitution
of ansatz (\ref{MM}) in Lagrangian (\ref{L2D}) yields
\begin{equation}
L_{\mathrm{eff}}=\mu N-\frac{\pi \Gamma \left( 2+\alpha /2\right) }{\left(
2\beta \right) ^{2-\alpha /2}}\left( 2\beta A_{1}^{2}+A_{2}^{2}\right)
+(1+\gamma )\left( \frac{\pi A_{1}^{4}}{4\beta }+\frac{\pi A_{2}^{4}}{%
32\beta ^{3}}\right) +\frac{\pi A_{1}^{2}A_{2}^{2}}{4\beta ^{2}}-\frac{%
2\lambda \pi A_{1}A_{2}}{\beta },  \label{va2}
\end{equation}%
cf. Eq. (\ref{va1}), where the total norm is $N=\pi \lbrack A_{1}^{2}/(\beta
)+A_{2}^{2}/(2\beta ^{2})]$, cf. Eq. (\ref{N2D}). The corresponding
variational equations are then derived as per Eq. (\ref{EL2D}).
\begin{figure}[h]
\begin{center}
\includegraphics[height=4.cm]{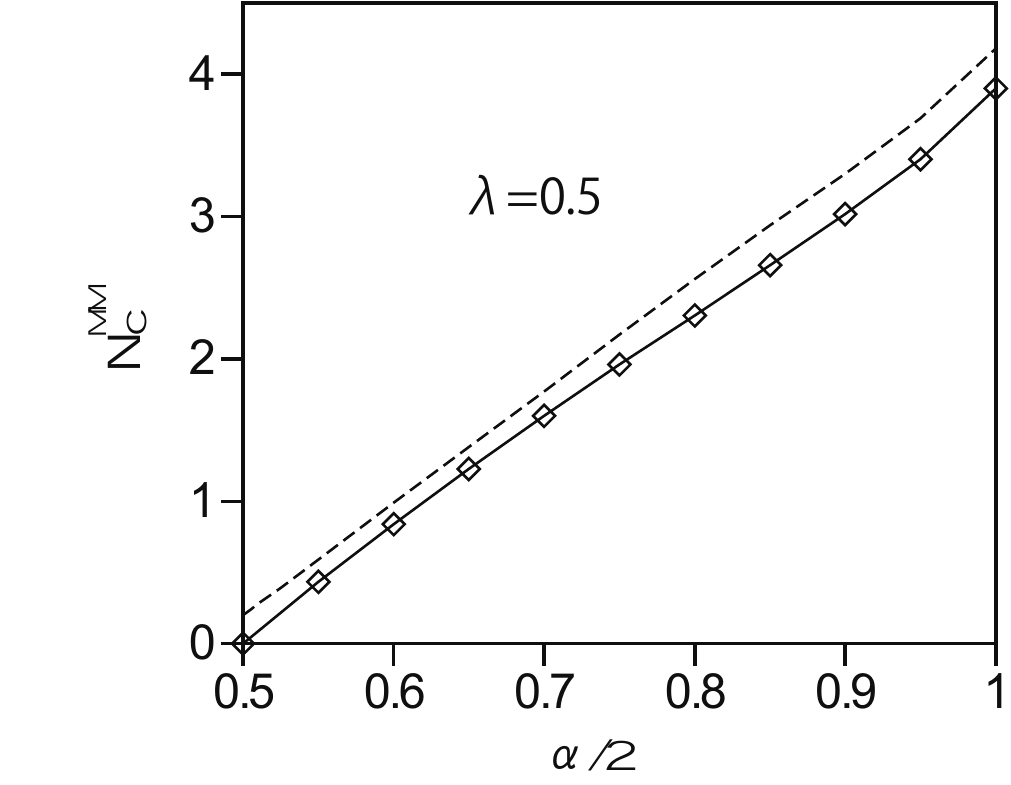}
\end{center}
\caption{The critical value of the norm of the MM (mixed-mode) solitons in
the 2D system, $N_{c}^{\mathrm{(MM)}}$, as a function of the half L\'{e}vy
index, $\protect\alpha /2$, for fixed $\protect\lambda =0.5$ and $\protect%
\gamma =2$ in Eq. (\protect\ref{2D}). At $1<\protect\alpha \leq 2$, the
system gives rise to stable MM solitons at $N<N_{c}^{\mathrm{(MM)}}$, and to
collapse at $N>N_{c}^{\mathrm{(MM)}}$. The chain of rhombuses and dashed
line represent, respectively, the numerical findings, obtained from the
imaginary-time simulations of Eq. (\protect\ref{2D}), and VA results
produced by Eq.~(\protect\ref{va2}).}
\label{fig6}
\end{figure}

Similar to the SV family, stable 2D solitons of the MM type exist at $\alpha
>1$, below the respective critical value of the norm, $N<N_{c}^{\mathrm{(MM)}%
}$. The numerically found and VA-predicted dependences $N_{c}^{\mathrm{(MM)}%
}(\alpha )$ are shown in Fig. \ref{fig6}, with $\lambda =0.5$ and $\gamma =2$
fixed in Eq. (\ref{2D}). Note that values of $N_{c}^{\mathrm{(MM)}}$ in Fig. %
\ref{fig6} are close to $2/3$ of the respective values of $N_{c}^{\mathrm{%
(SV)}}$, which are presented, for the same values of $\lambda $ and $\alpha $%
, in Fig. \ref{fig3}(a). This relation is explained by the same argument
which was presented, for $\alpha =2$, in Ref. \cite{we}: taking into regard
that the dominant contribution\ to the total norm of the SV soliton is
produced by the zero-vorticity component, and that the nonlinear interaction
between the MM components may be, roughly speaking, absorbed by means of
rescaling, $u_{\pm }\rightarrow \left( 1+\gamma \right) ^{-1/2}u_{\pm }$, an
approximate relation between the SV and MM norms is $N^{\mathrm{(MM)}%
}\approx 2\left( 1+\gamma \right) ^{-1}N^{\mathrm{(SV)}}$. For $\gamma =2$,
it amounts to the above-mentioned norm ratio, $N^{\mathrm{(MM)}}/N^{\mathrm{%
(SV)}}\approx 2/3$. A similar result, $N^{\mathrm{(MM)}}/N^{\mathrm{(SV)}%
}\approx 2/(1+\gamma )$, can be obtained for other values of $\gamma >1$,
which maintain the MMs as stable states.

\begin{figure}[h]
\begin{center}
\includegraphics[height=4.cm]{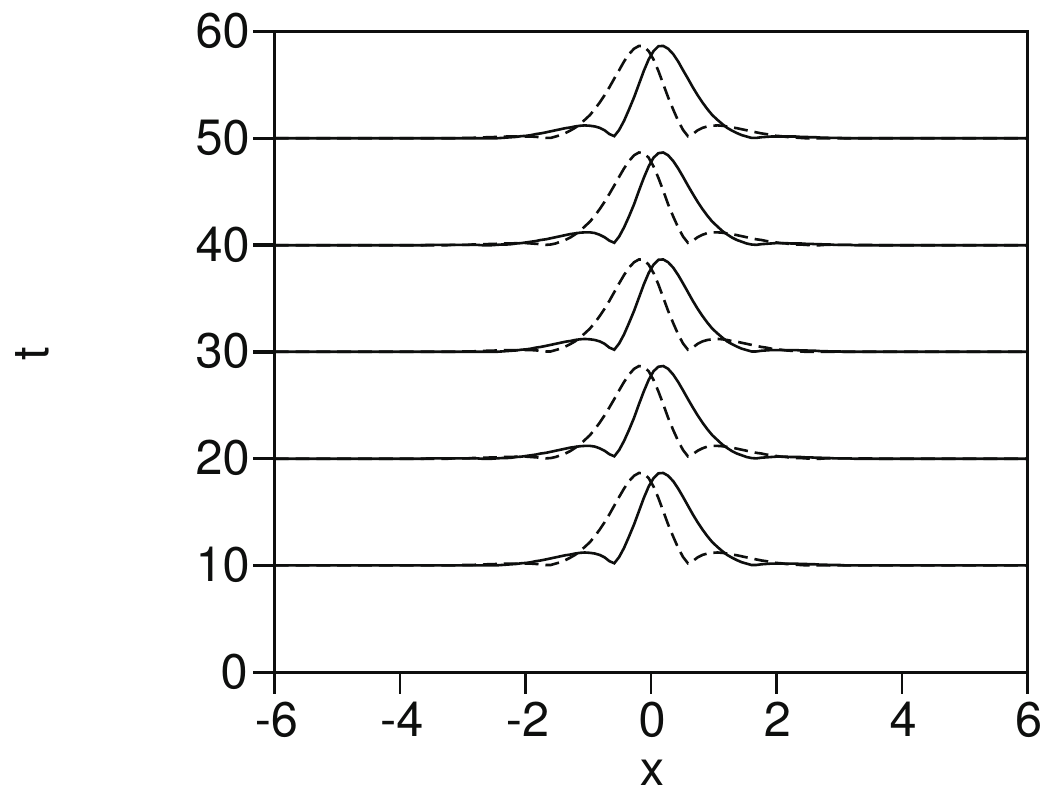}
\end{center}
\caption{The stable evolution of components $|\protect\phi _{+}(x,y)|$ and $|%
\protect\phi _{-}(x,y)|$ of a typical MM soliton, shown by solid and dashed
lines, respectively, in the cross section of $y=0$. The parameters are $%
\protect\lambda =1$, $\protect\alpha =1.5$, $\protect\gamma =2$. The total
norm of the soliton is $N=0.8$. The input was produced by simulations of Eq.
(\protect\ref{2D}) in imaginary time.}
\label{fig7}
\end{figure}

If $N<N_{c}^{\mathrm{(MM)}}$ is fixed, while the SOC strength $\lambda $ is
varying, the dependence of the MM's amplitude on $\lambda $ is similar to
that for the SVs plotted in Fig. \ref{fig3}(b), i.e., the amplitude vanishes
in the limit of $\lambda \rightarrow 0$ according to the same relation (\ref%
{scaling}) which is produced above for the SVs. Figure \ref{fig7} displays
the evolution of components $|\phi _{\pm }(x,y)|$ at $\lambda =1$, $\alpha
=1.5$, $\gamma =2$ and $N=0.8$, as produced by real-time simulations of Eq. (%
\ref{2D}) (note that these parameters correspond to a state located beneath
the stability boundary in Fig. \ref{fig6}). The plots clearly demonstrate
the structure and stability of the generic MM.

At $N>N_{c}^{\mathrm{(MM)}}$, the MM-shaped input leads to the collapse. In
the range of $\alpha <1$, where SV solitons do not exist, MM solitons were
not found either.

\section{Conclusion}

This works aims to consider effects of SOC (spin-orbit coupling) in 1D and
2D binary matter waves with the fractional kinetic-energy operator and the
usual cubic self- and cross-attractive nonlinearity. This is a model of BEC
composed of particles moving by L\'{e}vy flights, characterized by the value
of the L\'{e}vy index, $\alpha <2$ (the normal, non-fractional, kinetic
energy corresponds to $\alpha =2$). In the 1D setting, the effect of the SOC
is not dramatic, leading to decrease of the norm at which the collapse takes
place in the system with $\alpha =1$. Essential effects are predicted in the
2D setting. In that case, the SOC creates regions of stable solitons of the
SV (semi-vortex) and MM\ (mixed-mode) solitons in the interval of $1<\alpha
<2$, where the supercritical collapse occurs and no stable modes exist in
the absence of SOC. The stable solitons exist at values of the norm below
the respective critical values, $N<N_{c}^{\mathrm{(SV,MM)}}$. Amplitudes of
the stable solitons in these regions vanish along with SOC strength, $%
\lambda $, as per Eq. (\ref{scaling}).

As an extension of the present analysis, it may be relevant to consider
moving solitons and collisions between them. The 2D system may also be used
to predict other patterns, such as vortex lattices, cf. Refs. \cite%
{vort-latt}-\cite{Fukuoka}.

\section*{Acknowledgment}

The work of B.A.M. was supported, in part, by the Israel Science Foundation
through grant No. 1286/17.

\end{document}